\documentclass[manuscript,screen, acmsmall]{acmart}

\AtBeginDocument{%
  }

\setcopyright{none}
\acmDOI{}
\acmISBN{}
\acmYear{2026}

\acmConference[]{[Under Review]}

\begin{document}

\title[InquiryBits: Sharing AI Conversation Traces]{InquiryBits: Sharing AI Conversation Traces to Support Collaboration Within Trust Boundaries}

\author{Caitlin Morris}
\email{camorris@media.mit.edu}
\orcid{0003-3883-2702}
\author{Pattie Maes}
\email{pattie@media.mit.edu}
\affiliation{%
  \institution{MIT Media Lab}
  \city{Cambridge}
  \state{Massachusetts}
  \country{USA}
}

\renewcommand{\shortauthors}{Morris et al.}

\begin{abstract}
AI chat tools are shifting problem-solving and brainstorming conversations away from colleagues and into private AI interactions, reducing the shared awareness that supports team coordination. We introduce InquiryBits, a system that shares minimal summaries of AI conversations within configurable trust boundaries, separating AI-only analysis from human-visible sharing. In a study with 80 professionals, we find that people are broadly willing to share these traces to support collaboration and avoid duplicating work - but only within bounded groups. Comfort drops sharply as audience expands beyond close teams; the level of detail shared matters less than who can see it, with a preference for more detail over less within trusted groups. These findings suggest that trust boundaries, more than information granularity, may be the most impactful design parameter.
\end{abstract}

\ccsdesc[500]{Human-centered computing~Collaborative and social computing systems and tools}
\ccsdesc[500]{Human-centered computing~Empirical studies in collaborative and social computing}
\ccsdesc[300]{Security and privacy~Human and societal aspects of security and privacy}
\ccsdesc[300]{Human-centered computing~HCI design and evaluation methods}
\ccsdesc[300]{Human-centered computing~Computer supported cooperative work}

\keywords{awareness systems, AI-mediated collaboration, knowledge sharing, workplace coordination}


\maketitle

\section{Introduction}
Collaborative teams rarely coordinate by sharing fully formed work. More often, coordination happens through process fragments: quick questions, half-formed ideas, problems encountered along the way. Over time, these fragments build a working model of who knows what, enabling members to draw on expertise beyond their own \cite{Woolley2024-by,Wegner1987-vf}. AI systems are now changing how these fragments develop and circulate. Increasingly, questions that would have been asked among peers and colleagues are queried privately through tools like ChatGPT or Claude \cite{huang2025aiwork}. The individual benefits are apparent: this process is faster and sometimes more reliable. However, in this shift, teams lose some of the informal visibility that helps support distributed cognition.

This creates an opportunity as much as a problem. AI conversations produce rich traces of cognitive process, from approaches and confusions to solutions. These traces could support coordination if shared appropriately. Two teammates independently working through similar problems could save significant time by discovering each other's progress \cite{Ackerman1998-kd}. However, sharing one's thinking process is much more vulnerable and sensitive than sharing results. One participant in our study noted the concern of feeling "like a fraud that I've had to ask for help on something that someone else may perceive that I should know."

We introduce InquiryBits, a prototype that investigates this tension by exploring how small a shared representation can be while still enabling useful discovery. We test lightweight signals - topic labels and filtered synopses - that might allow discovery of collaboration opportunities without requiring full transparency. The system matches based on what people are actively working on, rather than declared expertise. When overlap is detected, users decide whether to share - and with whom - within configurable trust boundaries.

This paper makes three contributions. First, it introduces a tiered visibility architecture that separates AI-only analysis from human-visible sharing. Second, it provides empirical evidence from a study with 80 professionals showing that trust boundaries may matter more than information granularity in determining sharing comfort. Third, it identifies design implications for AI-mediated awareness systems that aim to support coordination without creating a sense of surveillance.

\section{Related Work}
\paragraph{Awareness and Social Translucence.} Research in CSCW has long emphasized that coordination depends on visibility, but not necessarily on full transparency. Early work on shared workspaces showed that people rely on small, often peripheral cues to stay aligned with each other’s activity, rather than complete information \cite{Dourish1992-co}. Later work on social translucence made a similar point more directly: systems are most effective when they make certain aspects of activity visible while leaving others in the background \cite{Erickson2000-ds}.

\paragraph{Process Visibility in Collaborative Systems.} Many approaches for improving collaborative visibility assume that the activity of interest is already happening in a shared space, such as edits to a document or messages in a channel. InquiryBits stems from a different premise: interaction with an AI system usually happens in private, often closer to a record of thinking than a record of work. Work on expertise location provides one useful comparison. Expertise location systems typically capture relatively stable signals by drawing on profiles, past contributions, or communication patterns \cite{Ackerman2013-fe}. In contrast, AI conversations reflect what someone is dealing with in the moment of active work, often before it appears in any shared artifact.

Related work has also explored how traces of individual information-seeking activity can support collaborative work, particularly in situations where existing tools provide limited visibility into what others have already explored \cite{Komlodi2008-rb}. More broadly, collaborative search research has examined how people coordinate around shared information-seeking tasks through partial traces of search activity \cite{Morris2013-tq}. AI conversations differ from search histories in important ways, however. Rather than primarily capturing retrieval behavior, they often contain extended records of exploratory reasoning, uncertainty, and iterative problem-solving.

\paragraph{Sharing AI Artifacts.} Prior work has also explored sharing AI artifacts in collaborative spaces. FlowGPT \cite{Li2024-ox} studies platforms where users publicly share prompts and chatbots, but focuses on polished artifacts for community reuse, not in-progress thinking within trust networks. PaperPing \cite{Wang2025-hx} uses social context to recommend papers to research groups: close to our logic, but sharing external content rather than users' own cognitive process. Tools like OpenClaw enable autonomous agent-to-agent collaboration \cite{OpenClaw-Contributors2026-ao, Weidener2026-cj}, but operate in environments for agent autonomy rather than human collaboration. We instead explore whether minimal representations of private AI conversations can support match-triggered discovery between collaborators.

\section{System Design}

The system separates AI-only analysis from human-visible sharing through a three-layer architecture:
\begin{enumerate}
    \item \emph{Layer 1: Topic Signature.} Conversations are processed locally to extract topic labels and embeddings used for overlap detection within a user’s configured trust circle. No conversation content is visible to other users at this stage.
    \item \emph{Layer 2: Sharing Prompt.} When overlap is detected, both users are prompted to share a minimal synopsis of their conversation, such as a topic label or short summary. If both consent, contact information and selected details are exchanged.
    \item \emph{Layer 3: AI Context Bridge.} Users can optionally allow AI systems to exchange additional conversation context directly, enabling more informed assistance without exposing the underlying conversation to another person.
\end{enumerate}

Users configure which social scopes are eligible for matching, ranging from close collaborators to department- or organization-level visibility. The prototype assumes an initial consent step for match-seeking, with users choosing chat-by-chat which conversations are eligible for topic extraction and overlap detection rather than granting blanket access to all of their AI chats.

\begin{figure}[htbp]
  \centering
  \includegraphics[width=0.95\linewidth]{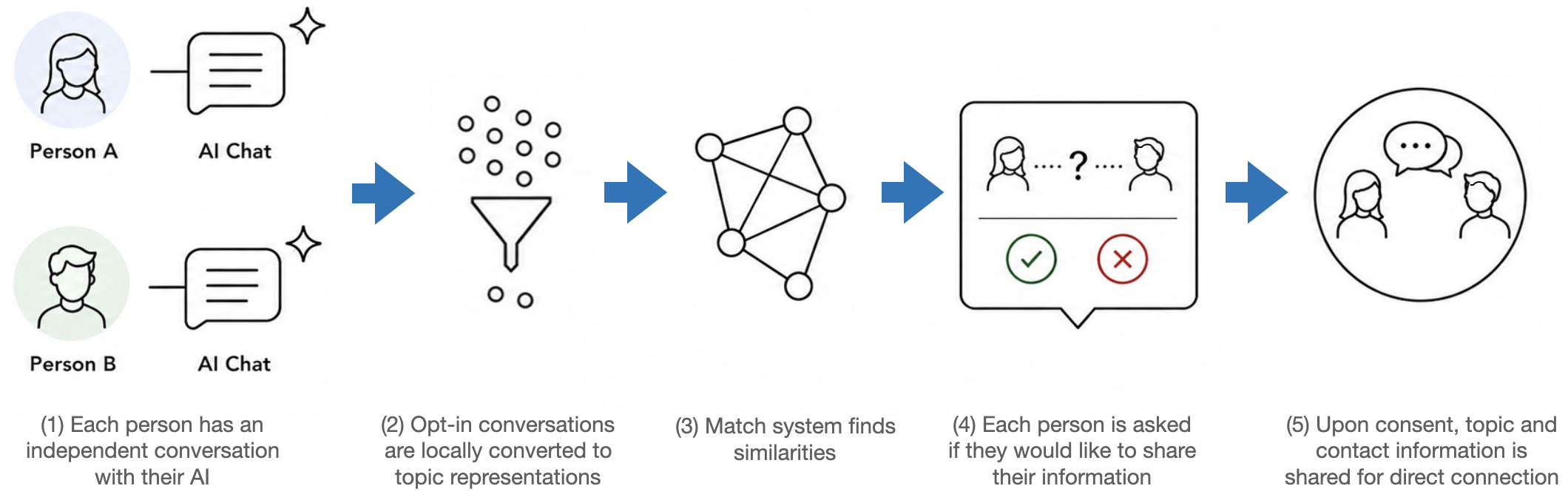}
  \caption{InquiryBits detects overlap between private AI conversations within configurable trust boundaries. When a potential match is found, users can choose whether to share a minimal synopsis of their conversation. If both users consent, synopsis details and contact information are exchanged. Conversations are locally processed for topic metadata; full chat content is not centrally visible.}
  \label{fig-system-overview}
  \Description{Diagram: InquiryBits detects overlap between private AI conversations within configurable trust boundaries. When a potential match is found, users can choose whether to share a minimal synopsis of their conversation. If both users consent, synopsis details and contact information are exchanged. Conversations are locally processed for topic metadata; full chat content is not centrally visible.}
\end{figure}

\section{Study Design}

We conducted a within-subjects study (N=80) using an interactive prototype to evaluate willingness to share and preferences across design parameters. The study was pre-registered and IRB-approved. Participants (50\% male, 48\% female; M age=35.0, SD=9.6) were recruited through Prolific and filtered for AI use at work: 60\% used AI chat tools daily and 40\% at least weekly. Work arrangements included fully remote (46\%), hybrid (36\%), and in-person (18\%) settings, across a range of team sizes and domains in both industry and academia.

\begin{figure}[htbp]
  \centering
  \includegraphics[width=0.7\linewidth]{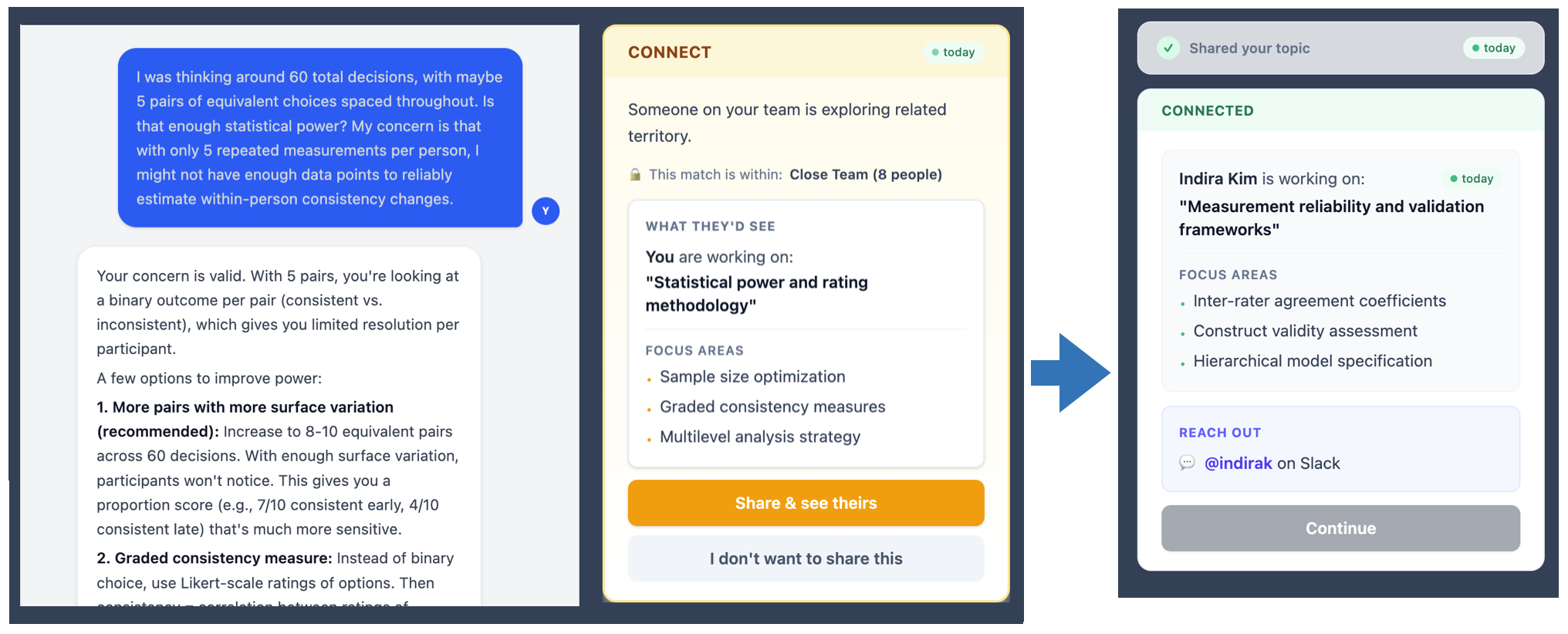}
  \caption{Prototype interface showing a match-triggered sharing prompt. Users are notified that someone within their selected trust level has had a related AI conversation, then review a generated summary of their own chat and choose whether to share it. If both consent, areas of overlap and contact information are revealed.}
  \label{fig-system-overview}
  \Description{Screen capture images of a prototype interface showing a match-triggered sharing prompt. Users can review a generated synopsis, see areas of overlap, and choose whether to share their InquiryBit with a colleague.}
\end{figure}

Participants were asked to evaluate the system by sharing a real work-related conversation they recently had with an AI chat (ChatGPT). Their conversation was imported to the interactive interface to provide a realistic example of a match-triggered sharing moment in their work. The conversation was processed locally in the participant’s browser and was not shared with the researchers or with other participants. Participants were informed that no one else would see their chat; an LLM (Claude Sonnet 4) was used only to identify possible moments for a simulated connection.

Each participant experienced two counterbalanced conditions: one showing only the topic label, one showing topic plus an expanded discussion synopsis. After viewing each condition, participants rated awareness value and connection likelihood, then decided whether to share. Additional measures assessed comfort across audience scopes, architecture preferences, and perceived value. Information granularity was tested through counterbalanced prototype conditions, while audience scope was tested through repeated comfort ratings across possible visibility groups.

\section{Results}

\subsection{System Adoption Interest}

When asked whether they would use such a system if it existed today, 79\% expressed interest, but most attached conditions: 32\% said yes unconditionally, 47\% yes with conditions, 16\% maybe, and 5\% no. The conditions participants named centered on control and trust: control over what's shared and with whom, privacy and data protection, team buy-in from colleagues, platform trust, and reasonable cost. In the prototype itself, sharing rates were high (92.5\%), though this likely reflects curiosity about the system rather than real-world intent; the conditional adoption responses are probably more predictive of actual deployment behavior.

\subsection{Trust Boundaries Matter More Than Information Level}

The difference in level of detail between conditions had minimal effect on decisions. Sharing rates were similar across conditions, and a McNemar exact test confirmed this difference was not significant (p=.219). When asked their preference for a real system, responses were mixed: 48\% preferred topic plus summary, 18\% preferred topic only, and 33\% wanted to choose per situation.

Those preferring more detail emphasized that specificity enables better decisions about relevance: "The topic can be broader and sometimes might not be useful for me. But if I know the focus areas, I can see what exact areas are covered." Another noted: "Having more details allows me to make a decision on whether I want to reach out to my colleague or not." Those preferring less detail cited concerns about "sensitive information I don't realize is in the chats."

\begin{figure}[htbp]
  \centering
  \includegraphics[width=0.95\linewidth]{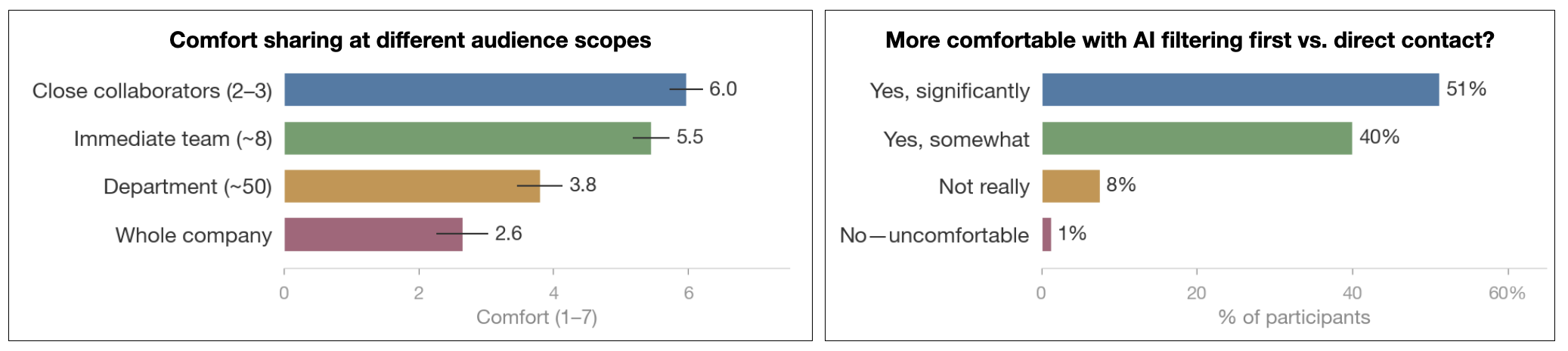}
  \caption{Left: comfort sharing AI conversation summaries across different audience scopes. Comfort drops sharply beyond close collaborators and immediate teams. Right: whether participants felt more comfortable with AI-mediated filtering before direct contact.}
  \label{fig-results-01}
  \Description{Two figures showing data: Left figure shows decreasing trust as scope of sharing expands from close team to broader company. Right figure shows greater comfort with AI-mediated sharing before direct contact.}
\end{figure}

In contrast, comfort varied sharply by audience scope (Fig. \ref{fig-results-01}, left). A Friedman test confirmed a significant effect across all four audience levels ($\chi$²(3) = 168.28, p < .001), with every pairwise drop significant after Bonferroni correction. Participants reported high comfort sharing with close collaborators (M = 5.97) and immediate team members (M = 5.45), but comfort fell substantially for department-level (M = 3.80) and company-wide sharing (M = 2.65). As one participant noted: "50 is a lot to assume you could actually vouch for."

\subsection{Architecture Preferences, Values and Concerns}
The system's separation between AI-only analysis and human-visible sharing increased comfort for most participants: The majority of participants indicated that AI-filtered connection, preserving privacy at the human-facing level, helped their comfort significantly or somewhat as compared to reaching out to a colleague directly, with only 9\% reporting little or no benefit (Fig. \ref{fig-results-01}, right). One participant conveyed their perspective: "Depending on what topic I choose, I might feel like some of the humans are judging me. But I don't think AI is judging me at all."

Participants also responded to a more speculative AI context bridge: an optional feature that would allow AI-to-AI sharing of more extensive chat details without direct human visibility. When participants opted to share, 87\% also enabled this feature. When comparing AI-to-AI sharing with direct connection to a colleague, 59\% found AI-to-AI sharing more comfortable, 22\% less comfortable, and 19\% about the same. However, transparency emerged as a critical concern. "Not knowing exactly what my AI is telling their AI" was a repeated theme. Participants wanted visibility and control: "Having a clear overview on what is being shared, and reminders to check whether I still want to share. Maybe even setting a limit on how long the 'share' lasts." These responses suggest that AI-to-AI sharing may feel less socially exposing than direct human disclosure, but that its acceptability depends on whether users can inspect and limit what is exchanged.

When explaining individual sharing decisions, participants most often cited collaboration potential (74\%) and efficiency (64\%). One noted: "The collaborative nature is extremely valuable to me. I think anything that removes friction from that process is a plus." But when asked to rank their single most important motivation, efficiency was the greatest driver: 48\% ranked "avoid duplicating work" first, compared to 21\% for finding collaborators. This suggests that while collaboration is broadly valued, efficiency may be the more reliable driver of adoption. 

Concerns centered on two themes. First, control: "I would be uncomfortable if I did not have clear control over what is being shared." Second, work-in-progress visibility: "The only thing that makes me uneasy is the level to which another person's AI might recall or remember my work. Especially if my work was sloppy because it was an early draft." Notably, some concerns were interpersonal rather than systemic: "It depends on the people I am working with, not necessarily the system itself."

\section{Discussion}

\subsection{Trust Boundaries and Control as Primary Design Parameters}
In this prototype context, who could see information mattered more for participant comfort than how much detail was shared. The relatively small differences between topic-only and topic-plus-summary conditions contrast sharply with the drop in comfort across audience scopes. Participants were broadly comfortable sharing within close teams, but substantially less so at department or company scale.
The adoption findings reinforce this point. Most participants expressed conditional willingness to use the system, with conditions centered on control over visibility boundaries rather than information detail. This tension between awareness and exposure echoes longstanding CSCW findings that systems designed to improve coordination can also create new forms of social and privacy concern \cite{Hudson1996-td}. Participants generally accepted AI-mediated matching and AI-to-AI context exchange, but repeatedly emphasized the need for transparency about what information would be shared and with whom.

\subsection{Design Implications}
These findings suggest several design implications for AI-mediated awareness systems. Systems should foreground clear and adjustable trust boundaries, provide preview mechanisms for AI-mediated sharing, and maintain visible separation between AI-only analysis and human-visible disclosure. Participants consistently framed the system in terms of practical coordination benefits, especially reducing duplicated work, suggesting that concrete efficiency gains may be a more compelling adoption framing than general desire for collaboration.

\subsection{Limitations}
This study uses a prototype probe to evaluate willingness to share and self-reported preferences; actual deployment behavior may differ. We note the limitations in ecological validity: participants evaluated a simulated match rather than real-time connection to a real colleague, in a context with no genuine professional stakes. This gap between prototype evaluation and organizational deployment is likely to affect both sharing willingness and the conditions people attach to adoption. The sample also skews toward heavy AI users by design, which may overrepresent comfort with AI-mediated sharing relative to broader professional populations.

\section{Conclusions}

AI chat tools are reshaping knowledge work by shifting process thinking into private, AI-mediated spaces. This reduces the visibility that supports coordination while generating new traces of in-progress work and reasoning. InquiryBits explores how these traces can be shared in a limited and controlled way to restore some of that lost visibility. The findings suggest that professionals are open to sharing minimal representations of AI conversations when visibility remains bounded and controllable.
The central design challenge is not how much detail to include, but how clearly and reliably trust boundaries are defined. Within bounded circles, there is appetite for sharing; beyond them, comfort drops sharply. Realizing this potential will depend on giving users genuine control over sharing and maintaining transparency about what AI communicates on their behalf.

\begin{acks}
This work was partially supported by funding from the MIT Morningside Academy for Design. We thank the participants for their feedback and perspectives. The authors used generative AI in the creation of this manuscript to suggest improvements for clarity and length. The authors take full responsibility for the content of this article.
\end{acks}

\bibliographystyle{ACM-Reference-Format}
\bibliography{paperpile}

@MISC{huang2025aiwork,
  title        = "How {AI} Is Transforming Work at Anthropic",
  author       = "{Huang, S., Seethor, B., Durmus, E., Handa, K., McCain, M.,
                  Stern, M., Ganguli, D.}",
  month        =  dec,
  year         =  2025,
  howpublished = "\url{https://anthropic.com/research/how-ai-is-transforming-work-at-anthropic/}",
  note         = "Accessed: 2026-5-6"
}

@ARTICLE{Erickson2000-ds,
  title     = "Social translucence: an approach to designing systems that
               support social processes",
  author    = "Erickson, Thomas and Kellogg, Wendy A",
  journal   = "ACM Trans. Comput. Hum. Interact.",
  publisher = "Association for Computing Machinery (ACM)",
  volume    =  7,
  number    =  1,
  pages     = "59--83",
  month     =  mar,
  year      =  2000,
  language  = "en"
}

@INCOLLECTION{Wegner1987-vf,
  title     = "Transactive memory: A contemporary analysis of the group mind",
  author    = "Wegner, Daniel M",
  booktitle = "Theories of Group Behavior",
  publisher = "Springer New York",
  address   = "New York, NY",
  pages     = "185--208",
  year      =  1987,
  language  = "en"
}

@INPROCEEDINGS{Li2024-ox,
  title     = "{FlowGPT}: Exploring domains, output modalities, and goals of
               community-generated {AI} chatbots",
  author    = "Li, Xian and Han, Yuanning and Liu, Di and An, Pengcheng and Niu,
               Shuo",
  booktitle = "Companion Publication of the 2024 Conference on
               Computer-Supported Cooperative Work and Social Computing",
  publisher = "ACM",
  address   = "New York, NY, USA",
  pages     = "355--361",
  month     =  nov,
  year      =  2024
}

@INPROCEEDINGS{Wang2025-hx,
  title     = "{PaperPing}: A socially-aware {AI} agent that recommends academic
               papers to research group chats with contextualized explanations",
  author    = "Wang, Ruotong and Zhou, Xinyi and Qiu, Lin and Chang, Joseph Chee
               and Bragg, Jonathan and Zhang, Amy X",
  booktitle = "Companion Publication of the 2025 Conference on
               Computer-Supported Cooperative Work and Social Computing",
  publisher = "ACM",
  address   = "New York, NY, USA",
  pages     = "532--535",
  month     =  oct,
  year      =  2025
}

@ARTICLE{Woolley2024-by,
  title     = "Understanding Collective Intelligence: Investigating the role of
               collective memory, attention, and reasoning processes",
  author    = "Woolley, Anita Williams and Gupta, Pranav",
  journal   = "Perspect. Psychol. Sci.",
  publisher = "SAGE Publications",
  volume    =  19,
  number    =  2,
  pages     = "344--354",
  month     =  mar,
  year      =  2024,
  language  = "en"
}

@ARTICLE{Ackerman2013-fe,
  title     = "Sharing knowledge and expertise: The {CSCW} view of knowledge
               management",
  author    = "Ackerman, Mark S and Dachtera, Juri and Pipek, Volkmar and Wulf,
               Volker",
  journal   = "Comput. Support. Coop. Work",
  publisher = "Springer Science and Business Media LLC",
  volume    =  22,
  number    = "4-6",
  pages     = "531--573",
  month     =  aug,
  year      =  2013,
  language  = "en"
}

@INPROCEEDINGS{Dourish1992-co,
  title     = "Awareness and coordination in shared workspaces",
  author    = "Dourish, Paul and Bellotti, Victoria",
  booktitle = "Proceedings of the 1992 ACM conference on Computer-supported
               cooperative work - CSCW '92",
  publisher = "ACM Press",
  address   = "New York, New York, USA",
  year      =  1992
}

@MISC{OpenClaw-Contributors2026-ao,
  title  = "{OpenClaw}: Your own personal {AI} assistant.
            https://github.com/openclaw/openclaw",
  author = "{OpenClaw Contributors}",
  year   =  2026
}

@ARTICLE{Weidener2026-cj,
  title         = "From agent-only social networks to autonomous scientific
                   research: Lessons from {OpenClaw} and Moltbook, and the
                   architecture of {ClawdLab} and beach.Science",
  author        = "Weidener, Lukas and Brkić, Marko and Lee, Phillip and
                   Karlsson, Martin and Noessler, Kevin and Kohlhaas, Paul",
  journal       = "arXiv [cs.AI]",
  month         =  mar,
  year          =  2026,
  archivePrefix = "arXiv",
  primaryClass  = "cs.AI"
}

@ARTICLE{Komlodi2008-rb,
  title     = "Collaborative use of individual search histories",
  author    = "Komlodi, Anita and Lutters, Wayne G",
  journal   = "Interact. Comput.",
  publisher = "Oxford University Press (OUP)",
  volume    =  20,
  number    =  1,
  pages     = "184--198",
  month     =  jan,
  year      =  2008,
  language  = "en"
}

@ARTICLE{Morris2013-tq,
  title    = "Collaborative search revisited",
  author   = "Morris, Meredith Ringel",
  journal  = "CSCW ’13",
  year     =  2013,
  language = "en"
}

@ARTICLE{Ackerman1998-kd,
  title     = "Augmenting organizational memory: a field study of answer garden",
  author    = "Ackerman, Mark S",
  journal   = "ACM Trans. Inf. Syst.",
  publisher = "Association for Computing Machinery (ACM)",
  volume    =  16,
  number    =  3,
  pages     = "203--224",
  month     =  jul,
  year      =  1998,
  language  = "en"
}

@INPROCEEDINGS{Hudson1996-td,
  title     = "Techniques for addressing fundamental privacy and disruption
               tradeoffs in awareness support systems",
  author    = "Hudson, Scott E and Smith, Ian",
  booktitle = "Proceedings of the 1996 ACM conference on Computer supported
               cooperative work - CSCW '96",
  publisher = "ACM Press",
  address   = "New York, New York, USA",
  year      =  1996
}

\end{document}